\documentclass[a4paper,5p,fleqn]{cas-dc}
\usepackage[numbers]{natbib}
\usepackage{multirow}
\usepackage[table]{xcolor}
\definecolor{lightgreen}{RGB}{144, 238, 144}
\definecolor{lightorange}{RGB}{255, 200, 100}
\definecolor{lightred}{RGB}{255, 100, 100}

\def\tsc#1{\csdef{#1}{\textsc{\lowercase{#1}}\xspace}}
\tsc{WGM}
\tsc{QE}
\tsc{EP}
\tsc{PMS}
\tsc{BEC}
\tsc{DE}

\begin{document}
\let\WriteBookmarks\relax
\def\floatpagepagefraction{1}
\def\textpagefraction{.001}
\shorttitle{Deep Learning Analysis of Prenatal Ultrasound}
\shortauthors{Y. Megahed et~al.}

\title [mode = title]{Deep Learning Analysis of Prenatal Ultrasound for Identification of Ventriculomegaly}                      



\author[1,2]{Youssef Megahed}[type=editor,
                        orcid=0009-0004-2595-5468]
                        
\cormark[1]
\ead{youssefmegahed@cmail.carleton.ca}

\credit{Conceptualization of this study, Methodology, Software}

\affiliation[1]{organization={Department of Systems and Computer Engineering, Carleton University}, 
                city={Ottawa},
                state={Ontario},
                country={Canada}}
                
\affiliation[2]{organization={Department of Methodological and Implementation Research, Ottawa Hospital Research Institute}, 
                city={Ottawa},
                state={Ontario},
                country={Canada}}

\affiliation[3]{organization={Department of Acute Care Research, Ottawa Hospital Research Institute}, 
                city={Ottawa},
                state={Ontario},
                country={Canada}}

\affiliation[4]{organization={Children's Hospital of Eastern Ontario Research Institute}, 
                city={Ottawa},
                state={Ontario},
                country={Canada}}

\affiliation[5]{organization={Better Outcomes Registry \& Network Ontario, Children’s Hospital of Eastern}, 
                city={Ottawa},
                state={Ontario},
                country={Canada}}

\affiliation[6]{organization={Department of Obstetrics and Gynecology, University of Ottawa}, 
                city={Ottawa},
                state={Ontario},
                country={Canada}}

\affiliation[7]{organization={School of Epidemiology and Public Health, University of Ottawa}, 
                city={Ottawa},
                state={Ontario},
                country={Canada}}
                
\affiliation[8]{organization={Department of Obstetrics, Gynecology \& Newborn Care, The Ottawa Hospital}, 
                city={Ottawa},
                state={Ontario},
                country={Canada}}

\affiliation[9]{organization={International and Global Health Office, University of Ottawa}, 
                city={Ottawa},
                state={Ontario},
                country={Canada}}

\affiliation[10]{organization={Department of Clinical Science and Translational Medicine, University of Ottawa}, 
                city={Ottawa},
                state={Ontario},
                country={Canada}}

\author[3]{Inok Lee}
\author[3]{Robin Ducharme}
\author[3,10]{Aylin Erman}
\author[3]{Olivier X. Miguel}
\author[4,5]{Kevin Dick}
\author[1]{Adrian D. C. Chan}
\author[1,2,4,7,10]{Steven Hawken}
\author[3,4,5,6,7,8,9]{Mark Walker}
\author[3,6,8]{Felipe Moretti}
\cormark[2]

\ead{fmoretti@toh.ca}


\credit{Data curation, Writing - Original draft preparation}


\cortext[cor1]{Corresponding author}
\cortext[cor2]{Principal corresponding author}

\begin{abstract}
The proposed study aimed to develop a deep learning model capable of detecting ventriculomegaly on prenatal ultrasound images. Ventriculomegaly is a prenatal condition characterized by dilated cerebral ventricles of the fetal brain and is important to diagnose early, as it can be associated with an increased risk for fetal aneuploidies and/or underlying genetic syndromes.
An Ultrasound Self-Supervised Foundation Model with Masked Autoencoding (USF-MAE), recently developed by our group, was fine-tuned for a binary classification task to distinguish fetal brain ultrasound images as either normal or showing ventriculomegaly. The USF-MAE incorporates a Vision Transformer encoder pretrained on more than 370,000 ultrasound images from the OpenUS-46 corpus. For this study, the pretrained encoder was adapted and fine-tuned on a curated dataset of fetal brain ultrasound images to optimize its performance for ventriculomegaly detection. Model evaluation was conducted using 5-fold cross-validation and an independent test cohort, and performance was quantified using accuracy, precision, recall, specificity, F1-score, and area under the receiver operating characteristic curve (AUC).
The proposed USF-MAE model reached an F1-score of 91.76\% on the 5-fold cross-validation and 91.78\% on the independent test set, with much higher F1-scores than those obtained by the baseline models by VGG-19 (72.39\% and 75.63\%), ResNet-50 (89.45\% and 89.22\%), and ViT-B/16 (86.73\% and 79.85\%), respectively. The model also showed a high mean test precision of 94.47\% and an accuracy of 97.24\%. Eigen-CAM (Eigen Class Activation Map) heatmaps showed that the model was focusing on the ventricle area for the diagnosis of ventriculomegaly, which has explainability and clinical plausibility.
The study showed that domain-specific pretraining on a large corpus of ultrasound images can lead to a superior classification performance and generalization to fetal brain anomaly detection. The proposed USF-MAE deep learning framework is a robust and explainable foundation for the reliable assessment of ventriculomegaly on prenatal ultrasound. 
\end{abstract}





\maketitle

\section{Introduction}
Deep Learning (DL) is a type of Machine Learning (ML) that uses artificial neural networks to automatically learn hierarchical feature representations from data, enabling a wide range of tasks such as prediction, classification, generation, and reconstruction. Due to its exceptional pattern recognition and its ability to process large amounts of data, DL is increasingly used in medicine for medical image processing. For example, DL is used for classification of images (e.g., normal or abnormal) to support diagnosis{~\hypersetup{hidelinks}\textcolor{blue}{\cite{b23,b24,b25,b26,b27}}}, object detection to annotate structures (e.g., tumour){~\hypersetup{hidelinks}\textcolor{blue}{\cite{b28,b29,b30}}}, and image segmentation to precisely contour structures{~\hypersetup{hidelinks}\textcolor{blue}{\cite{b1,b28,b31,b32,b33,b34}}}. It has already shown great success in diagnosing ophthalmologic conditions, breast cancer and lung nodules{~\hypersetup{hidelinks}\textcolor{blue}{\cite{b2}}}. Convolutional Neural Networks (CNN) and Vision Transformers (ViT) are the main DL architectures used for such applications{~\hypersetup{hidelinks}\textcolor{blue}{\cite{b3}}}.

\begin{figure*}
	\centering
	\includegraphics[width=.98\textwidth]{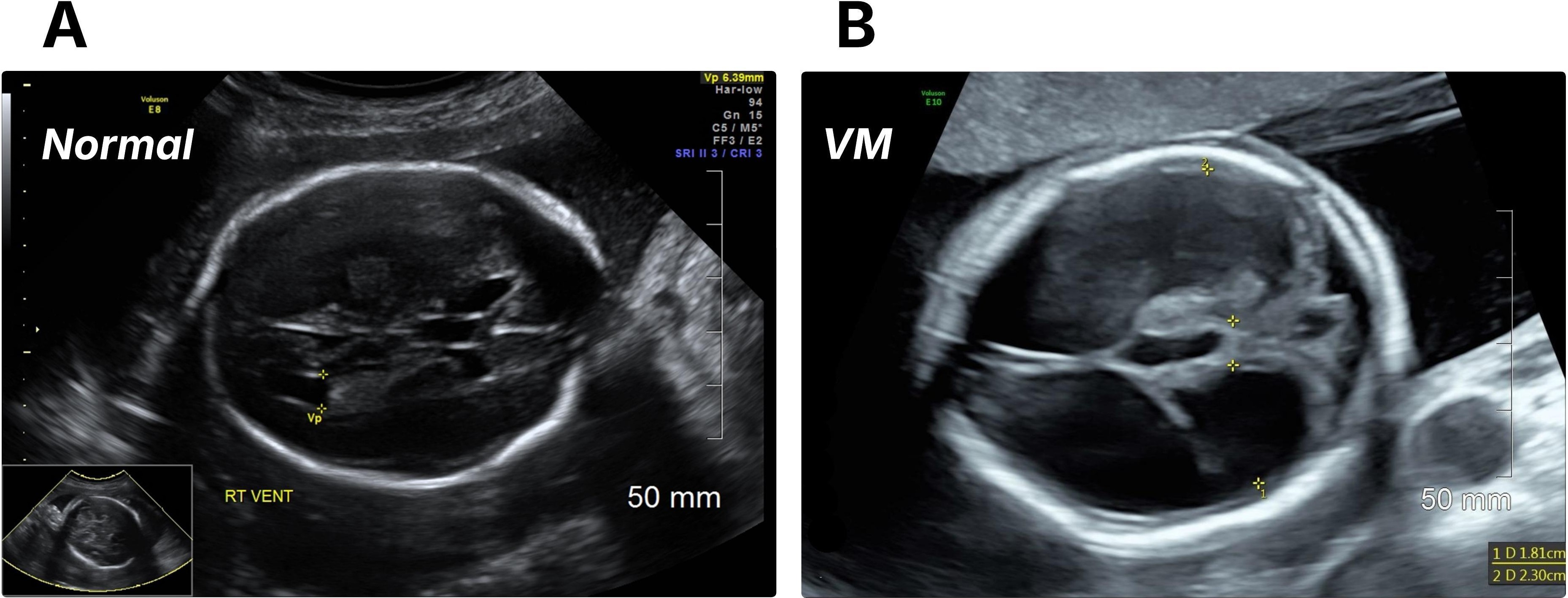}
	\caption{Prenatal ultrasound comparison of normal and ventriculomegaly (VM) fetal brains. (A) Normal fetal brain showing typical lateral ventricle size and configuration \textcolor{blue}{\cite{b35}}. (B) Fetal brain with VM demonstrating enlarged lateral ventricles.}
	\label{FIG:class_image_examples}
\end{figure*}

DL’s general success in medical image analysis has increased interest in its application to obstetric ultrasonography. It offers an exciting opportunity to enhance prenatal diagnosis. However, unlike other imaging modalities like CT and MRI, ultrasonography is very operator-dependent and thus has high inter-operator variability{~\hypersetup{hidelinks}\textcolor{blue}{\cite{b4}}}. This introduces additional noise to the image dataset, which might affect the accuracy and generalizability of the models{~\hypersetup{hidelinks}\textcolor{blue}{\cite{b5,b6}}}. Although several studies have explored the usefulness and performance of DL on fetal ultrasonography, continued investigation is needed to establish its clinical value.

Ventriculomegaly (VM) is defined as an enlargement of the lateral ventricles, greater than 10mm in the developing brain, and is associated with CNS and non-CNS malformations{~\hypersetup{hidelinks}\textcolor{blue}{\cite{b7}}}. Identification of this condition is first done on ultrasound, the first-line imaging modality used during prenatal screening. VM is also among the diagnoses with the highest false positive rate, which likely causes anxiety in parents, and may lead to unnecessary and potentially risky ensuing tests, highlighting the pressing need to improve diagnostic accuracy of VM on ultrasound{~\hypersetup{hidelinks}\textcolor{blue}{\cite{b7}}}. Other studies using DL for ventriculomegaly detection are either on CT or MRI{~\hypersetup{hidelinks}\textcolor{blue}{\cite{b8}}}, trained on relatively small datasets{~\hypersetup{hidelinks}\textcolor{blue}{\cite{b9}}}, measuring lateral ventricles without classification{~\hypersetup{hidelinks}\textcolor{blue}{\cite{b9}}}, or are not specific to VM{~\hypersetup{hidelinks}\textcolor{blue}{\cite{b10}}}.

Our research team has successfully applied DL to the ultrasound diagnosis of cystic hygroma and congenital anomalies of the kidneys and urinary tract{~\hypersetup{hidelinks}\textcolor{blue}{\cite{b11,b12}}}. Recently, we developed USF-MAE (Ultrasound Self-Supervised Foundation Model with Masked Autoencoding), the first large-scale, MAE foundation model pretrained exclusively on ultrasound data{~\hypersetup{hidelinks}\textcolor{blue}{\cite{b13}}}. USF-MAE demonstrated strong cross-anatomical generalization and label efficiency across multiple ultrasound classification tasks. Building upon this foundation, the present study applies USF-MAE to the prenatal ultrasound diagnosis of VM. Our contributions are twofold:

\begin{enumerate}
  \item To extend the investigation of foundation model–based deep learning approaches for ultrasound image analysis by leveraging a pretrained, self-supervised architecture specifically tailored to the ultrasound domain.
  \item To fine-tune this USF-MAE for the binary classification of fetal brain ultrasound images as either normal or showing VM.
\end{enumerate}

The performance of USF-MAE is further compared against common deep learning foundation models pretrained on natural images, such as convolutional and transformer-based architectures, which lack domain-specific ultrasound feature representations. This comparison highlights the benefits of modality-specific pretraining for achieving higher accuracy and generalization in fetal neuroimaging tasks.

\section{Methodology}
\subsection{Study Setting and Data Acquisition}
The study was approved by the Research Ethics Board. This retrospective study included ultrasound images from pregnant individuals with singleton or twin pregnancies obtained between June 2014 and May 2021 at a tertiary hospital in Eastern Ontario, Canada. All images were acquired using the GE Voluson V730, E08, and E10 ultrasound systems by professional obstetric sonographers and interpreted by maternal-fetal medicine specialists. Among these, we selected fetal brain images with or without VM that were obtained between 18 and 24 weeks of gestation ({\hypersetup{hidelinks}\textcolor{blue}{Fig.~\ref{FIG:class_image_examples}}} shows an example of each class). Multiple images were collected from patients who underwent several ultrasound examinations within the designated gestational age (GA) range.  

Images of two-dimensional (2D) transventricular (axial) plane at the level displaying the frontal horns and cavum septi pellucidi (CSP), ensuring symmetrical appearance of the cerebral hemispheres, were extracted from the institutional Picture Archiving and Communication System (PACS) and saved in Digital Imaging and Communication in Medicine (DICOM) format. VM cases were identified as having an atrial diameter of $\geq$ 10 mm on the transventricular plane{~\hypersetup{hidelinks}\textcolor{blue}{\cite{b22}}}. The images were classified into two groups, with or without VM, with no further classification based on severity. We excluded ultrasound images of fetuses with brain anomalies other than VM, as well as those not acquired in standard grayscale ultrasound format.   

\begin{table*}[htbp]
\centering
\caption{Dataset composition for VM classification. The dataset was divided into training, validation, and a held-out test set. The combined training+validation set (85\% of the total data) was further partitioned using 5-fold cross-validation for model development and hyperparameter tuning, while the test set (15\%) was kept completely separate and used only for final performance evaluation.}
\label{tab:us_datasets_distribution}
\setlength{\tabcolsep}{10pt}
\renewcommand{\arraystretch}{1.5}
\begin{tabular}{lccc}
\toprule
\textbf{Subset} & \textbf{Total (n, \%)} & \textbf{Normal (n, \%)} & \textbf{Ventriculomegaly (n, \%)} \\
\midrule
Training & 560 (68.04\%) & 462 (67.94\%) & 98 (68.53\%) \\
Validation & 140 (17.01\%) & 116 (17.06\%) & 24 (16.78\%) \\
Test (Hold-out) & 123 (14.95\%) & 102 (15.00\%) & 21 (14.69\%) \\
\textbf{Total} & \textbf{823 (100\%)} & \textbf{680 (100\%)} & \textbf{143 (100\%)} \\
\bottomrule
\end{tabular}
\end{table*}

\subsection{Dataset Distribution and Splitting Strategy}
The curated VM dataset contains 823 fetal brain ultrasound images, of which 680 are normal and 143 represent cases of VM ({\hypersetup{hidelinks}\textcolor{blue}{Table~\ref{tab:us_datasets_distribution}}}). This corresponds to about a 1:5 class imbalance ratio with the minority class being VM. The VM dataset was randomly stratified into training, validation, and independent held-out test sets. 560 of the images (68.04\%) in each fold were in the training set, 140 of the images (17.01\%) were in the validation set, and 123 of the images were in an independent hold-out test set, as summerized in{~\hypersetup{hidelinks}\textcolor{blue}{Table~\ref{tab:us_datasets_distribution}}}. Stratified sampling ensured that the class-wise distribution was also kept the same: approximately 68\% of both normal and VM cases are used for training, 17\% for validation, and about 15\% of them are kept aside as a final evaluation set. This level of stratification ensured that the majority and minority classes would be represented equally well through all phases of model development.

\subsection{Data Preprocessing}
Prior to training the DL models, a multi-stage preprocessing pipeline was applied to all ultrasound images to maintain patient privacy, remove visual artifacts, and yield a consistent dataset ready for fine-tuning the pretrained USF-MAE and baseline models. The entire workflow encompassed four major stages:(1) de-identification and scrubbing metadata; (2) removal of on-image annotations and graphical markers; (3) normalization plus resizing of images, and (4) controlled data augmentation that would assist in generalization for the model. Every step has been subjected to proper quality control that would ensure the diagnostically relevant features of the fetal brain are maintained while eliminating confusing visual information.

\textbf{1) De-identification and scrubbing:}
Raw ultrasound images frequently contain textual overlays displaying patient information such as name, hospital identification number, examination date, or operator initials. These overlays are typically located along the upper border of the scan. To remove such personal health information (PHI), we implemented an automated cropping procedure that removes a fixed vertical strip from the top portion of each image. The cropping height was empirically determined after visually inspecting a representative subset of scans to identify the region where textual metadata consistently appeared. All images were then visually checked to ensure that no identifiers remained visible.

This de-identification step ensures both institutional ethics and privacy rules compliance, as well as reducing irrelevant high-contrast text regions that may bias the model in feature learning. The scrubbing script maintained the original folder hierarchy, which will facilitate subsequent data management and reproducibility.

\begin{figure*}
	\centering
	\includegraphics[width=.98\textwidth]{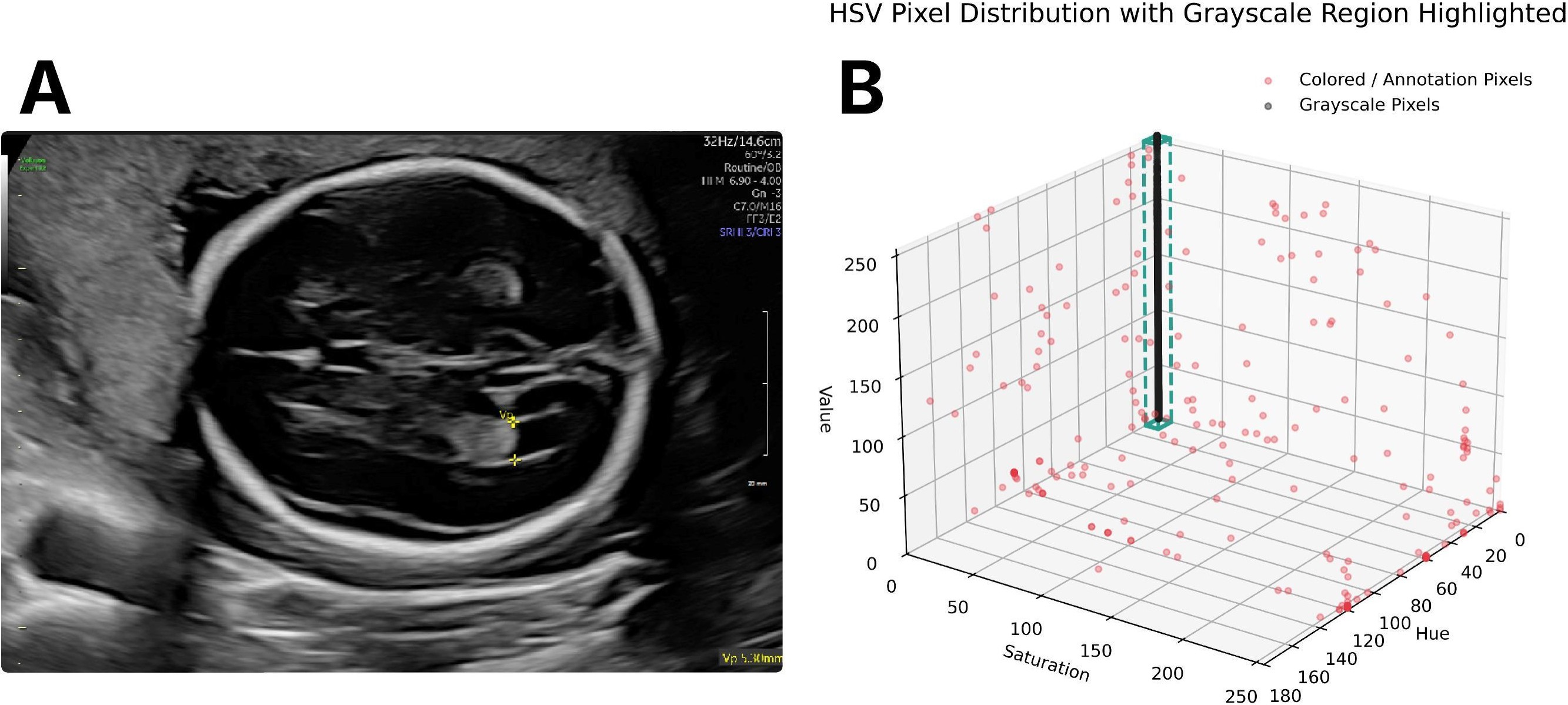}
	\caption{(A) Representative fetal ultrasound image used for analysis \textcolor{blue}{\cite{b35}}. (B) Corresponding pixel distribution mapped in HSV colour space, highlighting valid grayscale pixels versus those outside the grayscale range.}
	\label{FIG:pixel_distribution_figure}
\end{figure*}

\textbf{2) Annotation and marker removal:}
In the clinical practice of ultrasound, it is common for operators to annotate images right on the screen with measurement callipers, arrows, text labels or coloured highlights that identify anatomical landmarks, as can be seen in {\hypersetup{hidelinks}\textcolor{blue}{Fig.~\ref{FIG:pixel_distribution_figure}A}}. While helpful for human interpretation, these overlays introduce strong, non-biological edges and colour signals that may mislead a neural network to associate the markings rather than the anatomy with a specific class label. To mitigate this risk, we applied an automated annotation-removal algorithm similar to the pipeline previously described in our USF-MAE study{~\hypersetup{hidelinks}\textcolor{blue}{\cite{b13}}}.

Each RGB image was converted to the Hue-Saturation-Value (HSV) colour space, where colour components are more separable than in standard RGB representation, as shown in {\hypersetup{hidelinks}\textcolor{blue}{Fig.~\ref{FIG:pixel_distribution_figure}B}}. Regions exhibiting unusually high saturation or hue values, characteristic of synthetic markings such as yellow or green text, were detected using adaptive thresholding. The resulting binary mask was refined with morphological operations (dilation followed by closing) {\hypersetup{hidelinks}\textcolor{blue}{\cite{b15}}} to ensure that all annotation pixels and their immediate boundaries were included.

\begin{figure*}
	\centering
	\includegraphics[width=.98\textwidth]{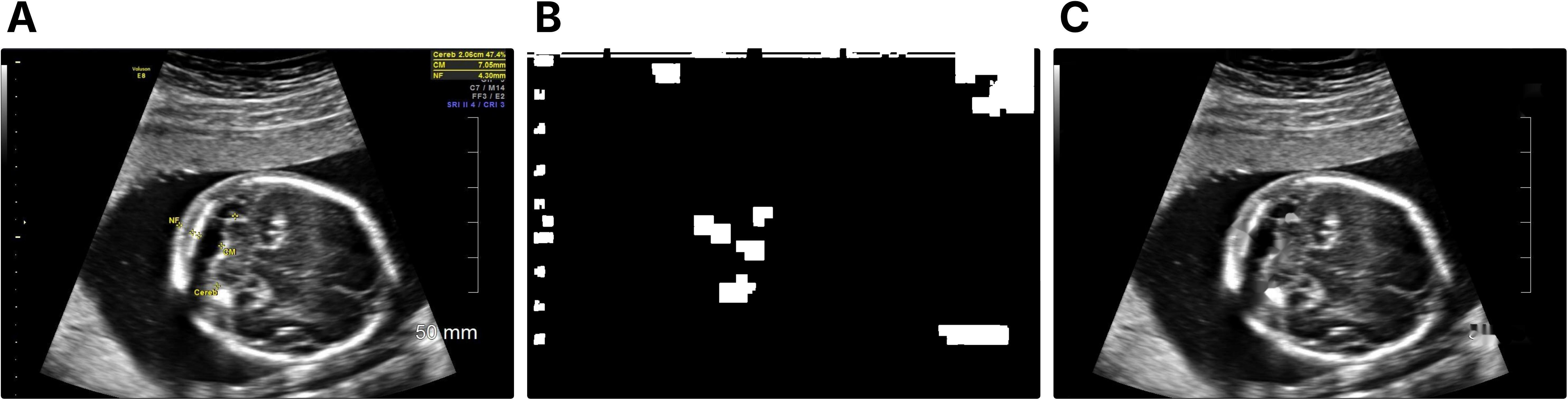}
	\caption{(A) Original fetal ultrasound image. (B) Binary mask identifying non-grayscale artifact regions. (C) Processed image after artifact removal and inpainting using a Navier-Stokes-based~\textcolor{blue}{\cite{b14}} inpainting method.}
	\label{FIG:preprocessing_pipeline_examples}
\end{figure*}

The missing areas were filled in with an inpainting technique based on Navier-Stokes equations{~\hypersetup{hidelinks}\textcolor{blue}{\cite{b14}}} ({\hypersetup{hidelinks}\textcolor{blue}{Fig.~\ref{FIG:preprocessing_pipeline_examples}C}}). This method is recognized for its capability to retain texture and intensity details from adjacent zones into the gaps. Following physical principles, it calculates the path of image gradients and prolongs isophote lines smoothly over the masked area, efficiently reconstructing essential echotexture designs. The inpainting made use of OpenCV’s version with a three-pixel radius since this provided a compromise between detailed texture completion and speed. Random groups of cleaned images were inspected by eye to ensure that no changes had occurred in the tissue structure, especially in fetal ventricles, and that no invented anomalies had been introduced (see {\hypersetup{hidelinks}\textcolor{blue}{Fig.~\ref{FIG:preprocessing_pipeline_examples}}}).

\textbf{3) Image normalization and resizing:}
After the de-annotation, the images were normalized to a uniform spatial resolution of 224$\times$224 pixels. This particular resolution was selected in order to make sure that it would be compatible with the ViT backbone being used in USF-MAE, and also as a trade-off between computational burden and enough structural detail for recognition of ventricular morphology. When resizing, we used nearest-neighbour interpolation to avoid generating artificial intermediate pixel values that could blur fine echogenic boundaries.

Pixel intensity values were normalized to match the statistical distribution of the ImageNet dataset (mean = [0.485, 0.456, 0.406]; standard deviation = [0.229, 0.224, 0.225]). Ultrasound images are different from general images used in ImageNet, but using the same normalization scheme enables transfer learning of pretrained models by matching the magnitudes of feature scales in the first layers of the network. This normalization reduces illumination variance as well and makes gradient updates more stable during optimization.

\textbf{4) Data augmentation:}
To enhance model generalization and to control overfitting, a controlled set of geometric data augmentations was applied during training. Since the appearance of anatomy in ultrasound is sensitive to the orientation of the probe, scaling, and pressure of tissue contact, we limited augmentations to transformations that reflect plausible variations which are encountered within clinical scanning rather than aggressive synthetic distortions. Specifically, each image was subjected to augmentations described as follows:

Random rotation of the images by an angle in the range 0$^{\circ}$ to 90$^{\circ}$ mimics variation of the probe angle between examinations. Flips horizontally and vertically with a probability of 0.5, corresponding to all possible orientations of acquisitions that naturally occur in fetal brain imaging. Random resized crop with scale between 0.5$\times$~and 2.0$\times$ the original field of view introduces modest zoom-in and zoom-out variations while keeping the aspect ratio.

Data augmentations were applied at runtime so that the model would see slightly different spatial presentations of the same anatomical pattern for every training epoch. No colour or brightness perturbations are applied to preserve grayscale intensity distributions typical in ultrasound images.

Preprocessing and augmentation were carried out with the help of PIL and OpenCV inside a custom PyTorch data loading pipeline. Therefore, this dataset, in its preprocessed version, was finally saved as PNG to keep the image quality intact while reducing file size and ensuring cross-compatibility between operating systems. The standardization procedures ensured getting a clean, standardized de-identified dataset that retains all the clinically relevant information for accurate classification of VM, preventing the model from learning spurious or patient-specific cues.

\subsection{Model Training Pipeline}
The VM classification model used the USF-MAE{~\hypersetup{hidelinks}\textcolor{blue}{\cite{b13}}} encoder fine-tuned on the curated fetal brain ultrasound dataset. The encoder was adapted to a binary classification objective for normal versus VM cases (i.e., presence or absence of VM). The training pipeline leveraged the rich representations specific to ultrasound learned during the large-scale self-supervised pretraining phase of USF-MAE. That entire flow included model initialization and fine-tuning with cross-validation, further optimized by adopting an adaptive training schedule. All experiments were run on standardized computational and software configurations to ensure reproducibility.

\begin{figure*}
	\centering
	\includegraphics[width=.93\textwidth]{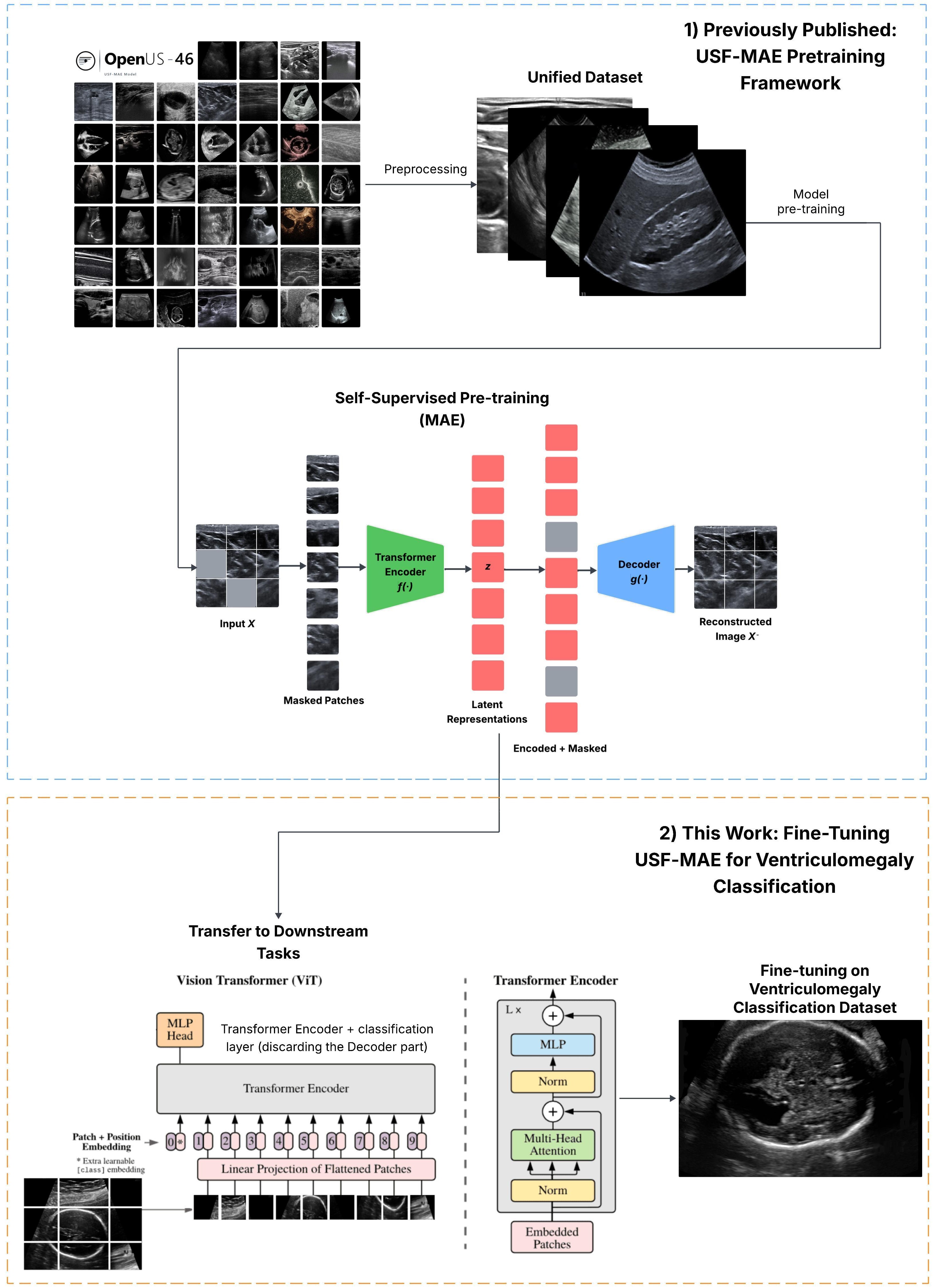}
	\caption{Overview of the proposed VM classification pipeline. (1) Previously published USF-MAE pretraining framework: A unified ultrasound dataset undergoes preprocessing and is used for self-supervised pretraining with a Masked Autoencoder (MAE), where random patches are masked (25\% masking ratio) and reconstructed to learn ultrasound-specific latent representations~\textcolor{blue}{\cite{b13}}. (2) This work: The pretrained encoder is transferred to the downstream VM classification task by discarding the MAE decoder and attaching a classification head. The model is then fine-tuned on the labeled VM dataset to enable clinical prediction.}
	\label{FIG:USF-MAE_training_arc_figure}
\end{figure*}

\textbf{1) USF-MAE Architecture:}
USF-MAE{~\hypersetup{hidelinks}\textcolor{blue}{\cite{b13}}} is a foundation model based on transformers that learns rich and generalizable feature representations from ultrasound data without manual labels. This architecture follows the principles of a Masked Autoencoder (MAE) first proposed by He et al. in 2022{~\hypersetup{hidelinks}\textcolor{blue}{\cite{b16}}}, but has been appropriately tuned for usage within the particular domain of ultrasound imaging. The USF-MAE pretraining and fine-tuning of its ViT encoder architectures are presented in {\hypersetup{hidelinks}\textcolor{blue}{Fig.~\ref{FIG:USF-MAE_training_arc_figure}}}.

The model employs a ViT backbone as the encoder and a lightweight decoder for image reconstruction. During pretraining, each ultrasound image is divided into fixed-size non-overlapping patches of 16$\times$16 pixels. A random subset of these patches, 25\% in our configuration, is masked, and the remaining visible patches are linearly projected into embedding vectors that serve as input tokens for the transformer encoder{~\hypersetup{hidelinks}\textcolor{blue}{\cite{b13}}}. The encoder processes only the visible tokens using multi-head self-attention layers to model spatial relationships among observed regions, thereby learning contextual feature representations that capture both local texture and global anatomical structure.

The decoder then receives a combination of the encoded visible tokens and learnable mask tokens representing the missing patches{~\hypersetup{hidelinks}\textcolor{blue}{\cite{b13}}}. With the help of its transformer blocks, the decoder reconstructs the image, including the masked parts, from contextual clues based on the visible areas. The model is pretrained by minimizing the Mean Squared Error (MSE) between the reconstructed image $\hat{x}$ and the original image $x$, encouraging the encoder to learn modality-specific priors inherent to ultrasound signals, such as speckle texture, acoustic shadowing, and tissue boundary patterns. The reconstruction loss is defined as:

\begin{equation}
    \mathcal{L}_{\text{MAE}} = \frac{1}{N} \sum_{i=1}^{N} \left\| \hat{x}_i - x_i \right\|_2^2,
\end{equation}
where $N$ is the number of pixels in the image, $x_i$ denotes the original pixel intensity, and $\hat{x}_i$ denotes the reconstructed pixel intensity.

USF-MAE was pretrained on $\sim$370,000 ultrasound images aggregated from 46 open-source datasets (OpenUS-46) covering 23 anatomical regions{~\hypersetup{hidelinks}\textcolor{blue}{\cite{b13}}} ({\hypersetup{hidelinks}\textcolor{blue}{Fig.~\ref{FIG:USF-MAE_training_arc_figure}}}). Such large-scale self-supervised pretraining helps build up within the encoder a holistic representation of ultrasound features over several organ systems; hence, a promising foundation to then fine-tune to new downstream tasks requiring specialization for diagnosis, such as the detection of VM in our study.

\textbf{2) Model initialization:}
The starting point for the experiments was the USF-MAE pretrained encoder, which had been previously trained on $\sim$370,000 ultrasound images from the OpenUS-46 corpus. The pretrained encoder provided a robust initialization for ultrasound image representation, allowing the model to generalize efficiently from unlabeled domain data to this specialized diagnostic task.

For this study, the MAE decoder module, responsible for reconstructing masked patches during pretraining, was removed and replaced with a task-specific classification head. The new head consisted of a fully connected (dense) layer mapping the encoder’s latent feature vector to two output nodes, corresponding to the binary classes: normal ventricles and VM. A SoftMax activation function was applied to convert logits into class probabilities. The rest of the encoder weights were retained and fine-tuned jointly with the classification head to enable end-to-end optimization.

\textbf{3) Fine-tuning process:}
To ensure unbiased evaluation and avoid data leakage, the dataset was divided into an independent hold-out test set and a training/validation set, as described in Section 2.2. The independent test set, consisting of 15\% of images from each class, was held out before model development and used only for final performance assessment. After completing the 5-fold cross-validation on the remaining 85\% of the data, each fold-specific checkpoint was applied to the independent held-out test set, and the resulting metrics were averaged across folds to obtain a robust estimate of test performance. This approach was adopted to enhance model generalization and mitigate the effects of class imbalance. The data set has a 1:5 ratio between VM and normal cases ({\hypersetup{hidelinks}\textcolor{blue}{Table~\ref{tab:us_datasets_distribution}}}). To ensure fair learning across both classes, the data (training and validation subsets) were randomly stratified into 5-folds, maintaining the same proportion of classes in each subset. Training was done on four out of five splits in every iteration (i.e., 80\% the dataset), with the remaining split being used as an internal validation set. The results obtained after running all splits were averaged out to get mean values for validation metrics.

Given the class imbalance, we applied class weighting within the cross-entropy loss function during model fine-tuning. Weights of classes were calculated inversely proportional to class frequencies in the training folds, thus effectively penalizing the model more for any misclassification on the positive class (VM). This leads to a weighting strategy that drives the model's focus more on minority classes during optimization steps and hence tunes sensitivity towards positive detections without losing total accuracy.

The preprocessing and augmentation pipeline, as described in Section 2.3 from the Methodology, was applied here in each fold. Data were loaded batch-wise and pre-processed on-the-fly with the help of PyTorch’s data loaders, thus ensuring consistent augmentations as well as memory efficiency.

\textbf{4) Optimization and hyperparameter configuration:}
The model was fine-tuned and optimized using AdamW{~\hypersetup{hidelinks}\textcolor{blue}{\cite{b17}}}, an optimizer that incorporates Adam's adaptive moment estimation with decoupled weight decay regularization. It is known to maintain stability and convergence efficiency during fine-tuning transformer-based architectures on medical image data.

An exhaustive grid search over batch size and learning rate within the below-stated ranges was conducted to try to identify the best training configuration: batch size $\in$ {32, 64, 128}, learning rate $\in$ {0.001, 0.0005, 0.0003, 0.00001}, and weight decay $\in$ {0.01, 0.05, 0.001, 0.0001}. A 5-fold cross-validation was run for every combination, and the one that produced the highest mean accuracy across folds was used. The optimal setup was found to be: learning rate = 0.0003, weight decay = 0.0001, and batch size = 128.

Cosine annealing with linear warm-up scheduled the learning rate such that it increased linearly in the first 10\% of total training epochs (warmup), then decayed smoothly following a half-cycle cosine curve. Such scheduling does not allow the model to converge prematurely and enables stable training at relatively high learning rates. Gradient clipping by maximum norm 1.0 was also applied as a stabilizing weight update and to avoid exploding gradients-common to neural networks.

Every training epoch comprised forward and backward passes over the data set and was thereafter followed by an evaluation step on the validation subset. The model's best state within each fold, validated by the highest possible validation accuracy, was saved as a checkpoint against degradation of performance due to overfitting. The resulting performance metrics were averaged across the best checkpoints from all the folds.

\textbf{5) Hardware and software configuration:}
Experiments ran in PyTorch. Hugging Face Transformers were used to load, train, and infer models. Data manipulation and evaluation used NumPy, Pandas, and scikit-learn.

Training was performed on a high-performance SLURM-managed GPU cluster at the hospital research institute. Each run utilized one NVIDIA L40S GPU with 128 GB used from the overall system's RAM. Each fold for every combination of hyperparameters took about 45 minutes at 100 epochs. Model checkpoints and logs were saved so that results could be tracked and reproduced across different experiments.

The USF-MAE foundation model weights and the OpenUS-46 pretraining corpus are publicly available and can be downloaded from our project repository (\href{https://github.com/Yusufii9/USF-MAE}{\textcolor{blue}{\textbf{GitHub Repository}}}: https://github.com/Yusufii9/USF-MAE). 

\subsection{Baseline Models for Comparison}
A benchmark was set through the selection of three well-known DL architectures as baselines. This included VGG-19{~\hypersetup{hidelinks}\textcolor{blue}{\cite{b18}}}, ResNet-50{~\hypersetup{hidelinks}\textcolor{blue}{\cite{b19}}}, and Vision Transformer Base (ViT-Base or ViT-B/16){~\hypersetup{hidelinks}\textcolor{blue}{\cite{b20}}}. These models are considered standard in the evolution of DL architectures, from convolutional feature extractors to transformer models capable of global attention modeling. They were also trained under the same preprocessing, data augmentation, and cross-validation protocols as used for the USF-MAE fine-tuning framework, so that direct comparison could be made.

\textbf{1) VGG-19:}
VGG-19{~\hypersetup{hidelinks}\textcolor{blue}{\cite{b18}}} is a type of DL model known for its depth and complexity. The VGG-19 model is composed of 3$\times$3 convolutional kernels padded with maximum-pooling layers and followed by two fully connected layers and one last classification layer representing some purely deep architecture with a fairly simple structure consisting of uniformly small-kernel convolutions that allow different levels of spatial features abstraction. For this study, it was initialized with pretrained weights from ImageNet-1K ($\sim$1.2 million training images over 1000 classes) fine-tuned end-to-end on the VM dataset using batch size 64, learning rate 0.0003, weight decay 0.05, the best hyperparameters found, trained for 100 epochs in order to return logits corresponding to the binary classes (normal vs. VM) after adapting the classification head of the model architecture appropriately. VGG-19 does not include any residual connections or attention mechanisms; however, the way it has such a deep convolutional stack offers strong baselines for evaluating classic feature hierarchies in ultrasound imagery.

\textbf{2) ResNet-50:}
The ResNet-50{~\hypersetup{hidelinks}\textcolor{blue}{\cite{b19}}} model implements residual learning using skip connections to effectively solve the problem of vanishing gradients in very deep networks. Each residual block learns a mapping which is an addition to its input, making it easier for the network to learn incremental changes rather than relearning everything from scratch. The architecture results in high representational capacity with efficient gradient propagation through layers. In this work, we have fine-tuned ResNet-50 that was initialized with ImageNet-1K pretrained weights using a batch size of 64, a learning rate of 0.001, and weight decay of 0.05 for 100 epochs. The final fully connected layer was updated such that it is replaced by a binary classifier specific to the VM detection task. The ResNet-50 provides convolutional baseline powerfulness and well-validation performance on numerous medical imaging applications.

\textbf{3) Vision Transformer Base (ViT-Base, or ViT-B/16):}
The ViT-B/16{~\hypersetup{hidelinks}\textcolor{blue}{\cite{b20}}} model represents the transformer paradigm for visual representation learning. Rather than using convolutions, it breaks every input image into 16$\times$16 non-overlapping patches linearly embedded as tokens added by positional encoding. These are fed through several layers of multi-head self-attention and feed-forward blocks: since dependencies at long distances and global context can be picked up by the model, it should be able to understand ultrasound images where structural cues are spread spatially across an image. Notably, ViT-B/16 shares the same backbone architecture as the encoder used in USF-MAE; the key difference lies in the pretraining strategy (ImageNet-1K supervised pretraining for ViT-B/16 versus large-scale self-supervised ultrasound-domain pretraining for USF-MAE). The ViT-B/16 model is initialized from ImageNet-1K pre-trained weights and used a batch size of 64, a learning rate equal to 0.0003 with weight decay set to 0.01 for 100 epochs. Its classification head was changed to provide two logits for the binary problem instead of the ImageNet-1K 1000 classes.

\textbf{4) Evaluation Conditions:}
Baseline models have been fine-tuned on the same 5-fold partition for cross-validation and held-out test set, and with the same preprocessing plus data augmentation steps described previously. The training epochs, the batch sizes, and the optimization schedules were kept alike in all experiments so that any differences seen would show the strength of each design rather than changes in the training method.

Together, these starting points offer a wide mix of design ideas, from known convolutional models to new transformer architectures, against the performance and generalization ability of the proposed USF-MAE architecture so that it can be critically evaluated.

\subsection{Performance Metrics}
The performance of the proposed USF-MAE model and all baseline architectures was quantified using several metrics that collectively characterize classification accuracy, diagnostic sensitivity, and discriminative power. Since this work is ultimately designed for a clinical application where verified detection of VM has important implications, those metrics which quantitatively describe the ability to correctly identify positive (VM) cases while simultaneously minimizing false negatives are emphasized. All metrics were computed on both the validation and independent held-out test sets, with final results being their mean values over the five cross-validation folds.

\textbf{1) Accuracy:}
Accuracy is defined as the total proportion of correctly classified samples out of all predictions made. It is defined as:
\begin{equation}
\text{Accuracy} = \frac{TP + TN}{TP + TN + FP + FN}
\end{equation}
where TP is True Positives, TN is True Negatives, FP is False Positives, and FN is False Negatives. Accuracy gives a general indicator of performance, but just accuracy alone cannot fully describe the quality of the model when applied to imbalanced datasets.

\textbf{2) Precision:}
Precision (Positive Predictive Value) is the ratio of true positive predictions to all positive predictions made by the model. A high precision value for this model would mean that it does not frequently label normal cases as having VM; hence, it is accurate when it makes a positive prediction, which is an important class-imbalance metric. It is expressed as:
\begin{equation}
\text{Precision} = \frac{TP}{TP + FP}
\end{equation}
Clinically, precision has great relevance in this situation because false-positive VM diagnoses lead to unwarranted parental anxiety and other unnecessary follow-up tests.

\textbf{3) Recall (Sensitivity):}
Recall or Sensitivity is defined as the proportion of actual positive cases that are correctly identified by the model. It is given by:
\begin{equation}
\text{Recall} = \frac{TP}{TP + FN}
\end{equation}
High recall means a strong ability to identify the real cases of VM with a low chance of missing any diagnosis. This is a major attribute in any clinical screening scenario.

\textbf{4) F1-Score:}
The F1-score is the harmonic mean of precision and recall, balancing the trade-off between the two. It is particularly informative for imbalanced datasets, where accuracy can be misleading. The F1-score is defined as:
\begin{equation}
\text{F1-score} = \frac {2\times (Precision \times Recall)}{Precision + Recall}
\end{equation}
A high F1-score speaks to strong general performance capturing both the model’s capacity to pick out real cases and its ability to avoid false alarms.

\textbf{5) Specificity:}
Specificity, or true negative rate, details the model's strength at rightly tagging normal (non-VM) cases. It is calculated as:
\begin{equation}
\text{Specificity} = \frac{TN}{TN + FP}
\end{equation}
Keeping specificity high is key when used in clinics since it lowers the chance of wrongly saying healthy fetuses have VM.

\textbf{6) The Area Under the ROC Curve, also known as AUC:}
The Area Under the Receiver Operating Characteristic Curve (AUC-ROC) also gives a threshold-independent measure of how well the model can differentiate between positive and negative classes. Recall at all thresholds plotted against 1-specificity gives the ROC curve, i.e. Recall (true positive rate) for all possible decision thresholds versus false positive rates, or 1-specificity. AUC values lie in the range [0, 1] where 0.5 is equivalent to random guessing and 1 is perfect discrimination. More specifically, higher AUC values mean that for most comparisons between a true VM case and a normal case, the model assigns higher scores to true VM cases. 

\begin{table*}[htbp]
\centering
\caption{Comparative performance of baseline and USF-MAE models on ventriculomegaly classification (Val Set).}
\label{tab:model_performance_val}
\setlength{\tabcolsep}{10pt}
\renewcommand{\arraystretch}{1.5}
\begin{tabular}{lcccccc}
\toprule

\textbf{Model} & \multicolumn{6}{c}{\textbf{Val Set (\%)}} \\
\cmidrule(lr){2-7}
 & \textbf{Accuracy} & \textbf{Precision} & \textbf{Recall} & \textbf{F1-score} & \textbf{Specificity} & \textbf{AUC} \\
\midrule

VGG-19 & 92.14 & 86.27 & 66.10 & 72.39 & 97.58 & 88.54 \\
ResNet-50 & 96.43 & 92.83 & 86.80 & 89.45 & 98.44 & 95.23 \\
ViT-B/16 & 95.57 & 90.94 & 83.57 & 86.73 & 98.09 & 95.72 \\
\textbf{USF-MAE (ours)} & \textbf{97.29} & \textbf{96.55} & \textbf{87.67} & \textbf{91.76} & \textbf{99.31} & \textbf{96.00} \\

\bottomrule
\end{tabular}
\end{table*}

\begin{table*}[htbp]
\centering
\caption{Comparative performance of baseline and USF-MAE models on ventriculomegaly classification (Test Set).}
\label{tab:model_performance_test}
\setlength{\tabcolsep}{10pt}
\renewcommand{\arraystretch}{1.5}

\begin{tabular}{lcccccc}
\toprule

\textbf{Model} & \multicolumn{6}{c}{\textbf{Test Set (\%)}} \\
\cmidrule(lr){2-7}
 & \textbf{Accuracy} & \textbf{Precision} & \textbf{Recall} & \textbf{F1-score} & \textbf{Specificity} & \textbf{AUC} \\
\midrule

VGG-19 & 92.68 & 84.70 & 69.52 & 75.63 & 97.45 & 88.43 \\
ResNet-50 & 96.26 & 88.25 & \textbf{90.48} & 89.22 & 97.45 & \textbf{97.50} \\
ViT-B/16 & 93.82 & 88.59 & 74.29 & 79.85 & 97.84 & 97.00 \\
\textbf{USF-MAE (ours)} & \textbf{97.24} & \textbf{94.47} & 89.52 & \textbf{91.78} & \textbf{98.82} & 97.00 \\
\bottomrule

\end{tabular}
\end{table*}

\section{Results}
{\hypersetup{hidelinks}\textcolor{blue}{Tables~\ref{tab:model_performance_val}}} and{~\hypersetup{hidelinks}\textcolor{blue}{\ref{tab:model_performance_test}}} report a summary of the comparative performance of all models evaluated on the cross-validation and the independent held-out test sets, respectively. The models considered in this comparison include the two convolutional baselines (VGG-19 and ResNet-50), the transformer baseline (ViT-B/16) as well as the proposed USF-MAE model fine-tuned for VM classification.

On the validation set ({\hypersetup{hidelinks}\textcolor{blue}{Table~\ref{tab:model_performance_val}}}), USF-MAE demonstrates the highest performance with respect to all the evaluation metrics, reaching an accuracy of 97.29\%, a precision of 96.55\%, a recall of 87.67\%, an F1-score of 91.76\%, a specificity of 99.31\%, and an AUC of 96.00\%. These results outperform the corresponding values of ViT-B/16 (accuracy 95.57\%, F1-score 86.73\%) and ResNet-50 (accuracy 96.43\%, F1-score 89.45\%) and demonstrate the utility of domain-specific pretraining on ultrasound data. While the observed validation AUC of 96.00\% points towards strong overall separability between classes, it needs to be interpreted with care in the context of the class imbalance in the dataset. The F1-score (91.76\%), the recall (87.67\%), and the precision (96.55\%) metrics, on the other hand, confirm that USF-MAE retains high detection rates of the minority (VM) class and is able to produce few false negatives on this dataset.

\begin{figure*}
	\centering
	\includegraphics[width=.98\textwidth]{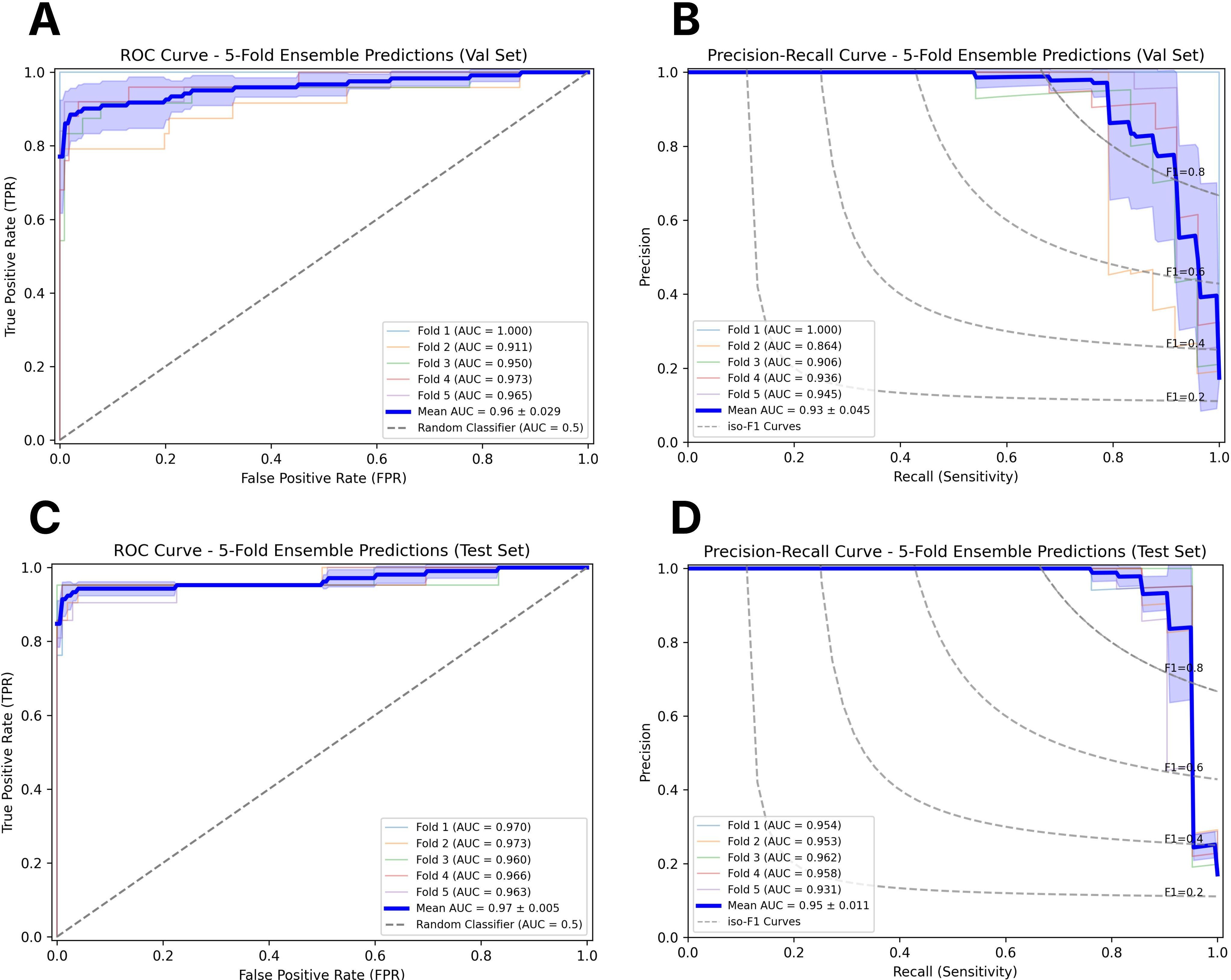}
	\caption{ROC (A, C) and Precision–Recall (B, D) curves for the 5-fold ensemble model on the validation (A, B) and test (C, D) sets. Curves for individual folds are shown along with the mean ensemble performance and variability across folds.}
	\label{FIG:val_and_test_curves}
\end{figure*}

The generalization of the model to an independent test set was evaluated by applying each of the 5 cross-validation checkpoints to the held-out test data and averaging the resulting performance metrics across folds ({\hypersetup{hidelinks}\textcolor{blue}{Table~\ref{tab:model_performance_test}}}). This cross-fold evaluation averaging provides a robust estimate of the model’s generalization capability. As can be seen, the proposed USF-MAE model retained its superior generalization under this evaluation protocol, yielding an accuracy of 97.24\%, a precision of 94.47\%, an F1-score of 91.78\%, and a specificity of 98.82\%, while maintaining high recall (89.52\%). Although the recall of ResNet-50 (90.48\%) was marginally higher, its overall precision and F1-score were inferior to those of USF-MAE, pointing towards a possible trade-off between true positive rate and confidence in predictions. The AUC of 97.00\% achieved by USF-MAE on the test set points towards the model's ability to robustly distinguish VM from normal cases under unseen clinical conditions.

Collectively, these results show that USF-MAE outperforms CNN and generic ViT baselines that were pretrained with ImageNet-1k weights for VM detection from fetal ultrasound images. The observed improvement is most likely due to its pretraining in a large-scale self-supervised learning task on the OpenUS-46 corpus{~\hypersetup{hidelinks}\textcolor{blue}{\cite{b13}}}, which results in the learning of ultrasound-specific textural and anatomical features that transfer well to the downstream task of prenatal imaging. The high precision and specificity of USF-MAE also point towards a reduction in false-positive diagnoses, which is a highly desirable property in a clinical setting where minimizing parental anxiety and reducing the burden of follow-up testing are important.

In summary, the results on the validation and test sets confirm that USF-MAE is a reliable and label-efficient foundation model for ultrasound-based fetal brain anomaly classification, achieving state-of-the-art performance even with limited labeled data.

\section{Discussion}
\subsection{Interpretation of Results and Model Performance}
Across both the validation set and the held-out independent test set, the USF-MAE model outperformed all baseline models, as indicated in {\hypersetup{hidelinks}\textcolor{blue}{Table~\ref{tab:model_performance_val}}} and {\hypersetup{hidelinks}\textcolor{blue}{\ref{tab:model_performance_test}}}. As shown in {\hypersetup{hidelinks}\textcolor{blue}{Fig.~\ref{FIG:val_and_test_curves}A-D}}, the ROC curves and Precision-Recall curves from the 5-fold ensemble had very narrow standard deviations, with the mean AUC values consistently greater than 0.96 for all 5 folds (held-out independent test set). This suggests that there was little variance in performance across the folds, and that the proposed model was able to generalize well despite the relatively small amount of data as well as the class imbalance between normal and VM cases.

On the held-out test set, the ROC curves in {\hypersetup{hidelinks}\textcolor{blue}{Fig.~\ref{FIG:val_and_test_curves}C}} demonstrated that almost all of the true VM cases were correctly classified, with a mean AUC of 0.97~$\pm$~0.005. The mean Precision-Recall curve in {\hypersetup{hidelinks}\textcolor{blue}{Fig.~\ref{FIG:val_and_test_curves}D}} had an area of 0.95~$\pm$~0.011, which showed that the model was able to maintain both high recall and high precision across all decision thresholds. These findings are aligned with clinical priorities, as recall was of particular importance to ensure that there were few missed true positive VM cases.

\begin{figure}
	\centering
	\includegraphics[width=.48\textwidth]{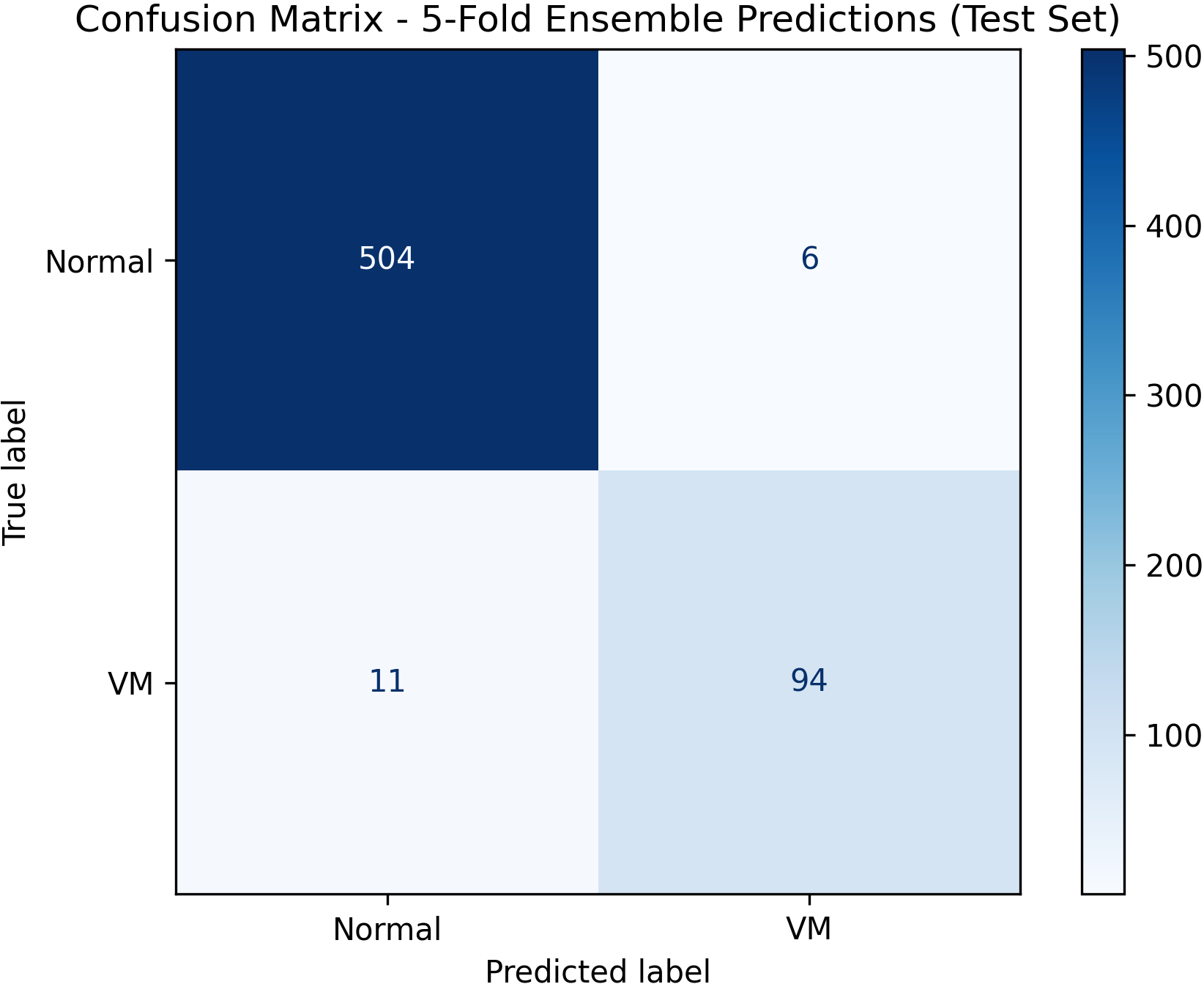}
	\caption{Confusion matrix for the 5-fold ensemble model on the test set, showing true versus predicted class labels for Normal and VM samples.}
	\label{FIG:model_testing_confusion_matrix}
\end{figure}

In addition, the confusion matrix in {\hypersetup{hidelinks}\textcolor{blue}{Fig.~\ref{FIG:model_testing_confusion_matrix}}} shows that there were only 17 misclassifications across the entire aggregated mean test cohorts (6 false positives and 11 false negatives). The model had a specificity of 98.8\% by correctly classifying 504 out of 510 normal cases, and a recall of 89.5\% by identifying 94 out of 105 VM cases. We considered this trade-off between recall and specificity to be appropriate for a prenatal screening setting, where overdiagnosis could lead to unnecessary parental anxiety but underdiagnosis could result in delayed further investigation of potential brain abnormalities. The robustness and consistency of these performance metrics across the validation and independent held-out test sets provided strong evidence to support the reliability of the USF-MAE encoder as a foundation model that could be applied to generalize to other unseen limited fetal ultrasound datasets.

These results overall demonstrated that the domain-specific pretraining on the large-scale OpenUS-46 corpus gave the proposed USF-MAE model a significant advantage over conventional CNNs and generic vision transformers. By pretraining the network to learn ultrasound-specific representations that captured both the global anatomical context and the fine-grained echotextural features, the model was able to achieve superior diagnostic performance using a minimal amount of labeled data. This highlights the potential of self-supervised foundation models to enable data-efficient clinical implementation in obstetric ultrasonography.

\subsection{Explainable Artificial Intelligence (XAI) Visualization Using Eigen-CAM}
\begin{figure*}
	\centering
	\includegraphics[width=.98\textwidth]{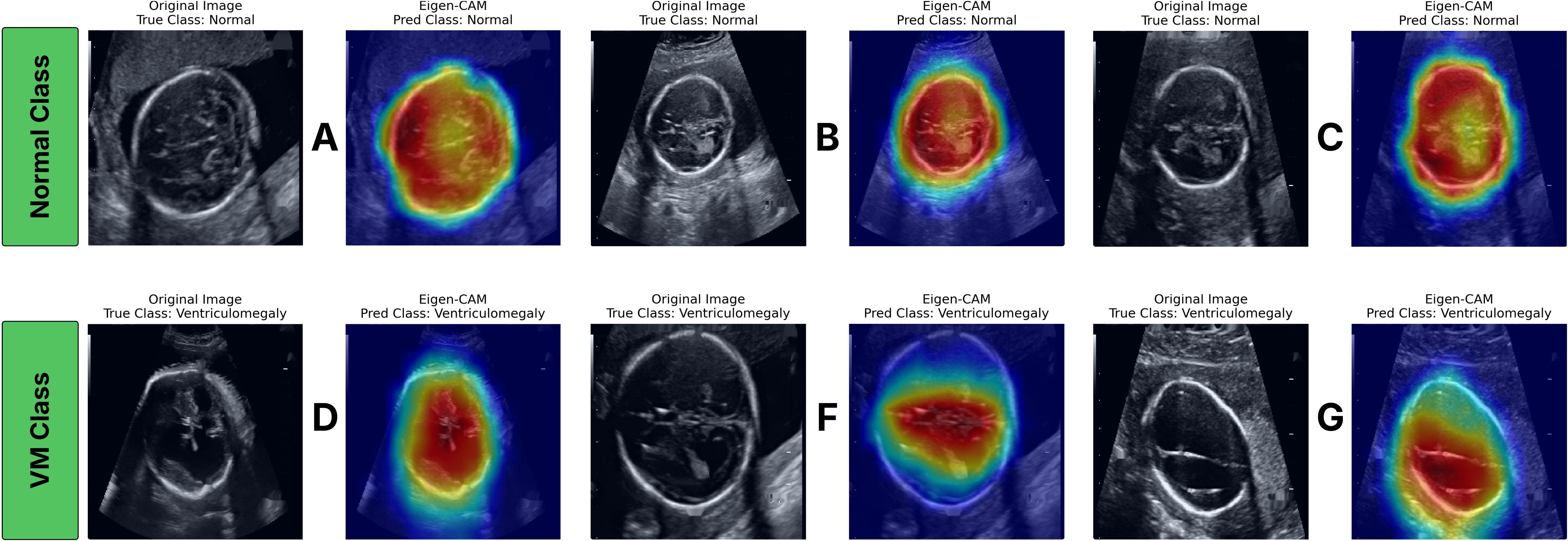}
	\caption{Representative ultrasound images and corresponding Eigen-CAM~\textcolor{blue}{\cite{b21}} heatmap visualizations for correctly classified cases from both classes. Columns show original images and the model's highlighted regions used for prediction. (A-C) Normal class examples. (D-G) Ventriculomegaly (VM) class examples.}
	\label{FIG:EigenCam_figure}
\end{figure*}

Explainability is a critical component in the clinical translation of DL models, particularly in high-stakes domains such as prenatal imaging. To interpret the decision-making process of the fine-tuned USF-MAE model, we employed the Eigen-CAM method{~\hypersetup{hidelinks}\textcolor{blue}{\cite{b21}}}, as visualized in {\hypersetup{hidelinks}\textcolor{blue}{Fig.~\ref{FIG:EigenCam_figure}}}. Eigen-CAM provides class activation maps (CAMs) based on the principal components of the model's activation space, highlighting the regions that most strongly influence its classification decisions without requiring gradient backpropagation.

In the top row of {\hypersetup{hidelinks}\textcolor{blue}{Fig.~\ref{FIG:EigenCam_figure} (A-C)}}, the model correctly focused on the periventricular region and the boundaries of the lateral ventricles for normal cases. In contrast, the bottom row (D-G) demonstrates that, for VM cases, the activation maps concentrated along the dilated ventricular cavities, with high-intensity regions corresponding to the expanded atrial width, the primary sonographic hallmark of VM. The smooth and spatially consistent activation distribution observed across all examples suggests that the model learned to recognize structural features relevant to fetal brain morphology rather than relying on spurious artifacts or background patterns.

While several gradient-based visualization techniques exist, including Grad-CAM and its variants, Eigen-CAM offers distinct advantages when applied to ViT-based architectures such as USF-MAE. Grad-CAM depends on spatial gradients propagated through convolutional feature maps to localize discriminative regions, a mechanism well-suited to CNNs but less effective for transformers, which operate on patch tokens without explicit convolutional kernels. In contrast, Eigen-CAM derives its visualizations directly from the dominant eigenvectors of the attention-based feature activations, capturing the principal modes of variance across all attention heads without gradient computation. This gradient-free formulation preserves the global context inherent to transformer attention mechanisms and avoids the instability or noise amplification that can arise during gradient backpropagation.

Consequently, Eigen-CAM produces smoother, anatomically coherent saliency maps that align closely with the true regions of clinical interest, in this case, the ventricular system. As shown in {\hypersetup{hidelinks}\textcolor{blue}{Fig.~\ref{FIG:EigenCam_figure}}}, the USF-MAE attention heatmaps unambiguously identify the ventricular edges in normal and VM cases. It is clear the model is making its predictions based on physiology-related features. This adds to model transparency and clinical trustworthiness, both of which will be important for eventual regulatory approval and real-world integration into obstetric workflows.

\subsection{Future Work and Clinical Implications}
Despite the robust diagnostic performance observed with the current model, there are several avenues for future work that can be explored to prepare for clinical integration. Expanding the training corpus with data from other institutions and ultrasound systems could be explored to evaluate the model's robustness across different devices, operators, and patient populations. Furthermore, incorporating 3D ultrasound volumes or cine-loop sequences of the fetal brain could be considered to allow for the learning of temporal features, which may be more sensitive to subtle changes in the ventricles throughout gestation.

In addition, semi-supervised or continual learning approaches can also be leveraged, taking advantage of the self-supervised nature of USF-MAE. As large amounts of unlabeled prenatal ultrasound data are stored in hospital archives, future versions of the model could be incrementally updated in the future to improve generalization without the need for manual annotations, while maintaining patient data privacy. On the clinical front, XAI visualizations such as Eigen-CAM could be embedded into reporting interfaces to provide real-time decision support, highlighting which regions of the ventricles contribute to the model's classification to improve interpretability for sonographers and fetal medicine specialists.

\section{Conclusion}
Our study proposed a DL framework utilizing the USF-MAE foundation model for the automated classification of VM in prenatal ultrasound images. The model employed ultrasound-specific representations learned from large-scale self-supervised pretraining and was subsequently fine-tuned on the fetal brain dataset. The performance of our framework was evaluated using 5-fold cross-validation and on independent test cohorts, and our results showed that our proposed approach achieved higher accuracy, precision, F1-score, and specificity in classifying VM in prenatal ultrasound images, compared to the baseline CNN and transformer models.

We can conclude that domain-specific pretraining on ultrasound data is beneficial for improving the generalization of features and label efficiency for downstream diagnostic tasks. The interpretability of the model was confirmed by Eigen-CAM heatmap visualizations, which revealed that the model's predictions were made based on relevant ventricular regions. These findings suggest that foundation models like USF-MAE can play an important role in enabling consistent and objective assessment of fetal brain structures on ultrasound, which may help in improving the diagnostic reliability of prenatal screening for VM and other related neurological conditions with further validation on multi-institutional and multi-system datasets.

\section*{Acknowledgments}
The authors thank Inbal Willner for her substantive work in assembling and organizing the ultrasound image dataset used in this study. Her contributions were essential to the successful completion of this research.

\section*{Declaration of competing interest}
The authors declare that they have no known competing financial interests or personal relationships that could have appeared to influence the work reported in this paper.


\end{document}